\documentclass{PoS}
\usepackage{amsmath}

\def\half{\frac{1}{2}}
\def\a{{\alpha}}
\newcommand{\eps}{\epsilon}
\def\IZ{\mathbb{Z}}

\def\IQ{\mathbb{Q}}

\newcommand{\hb}[1]{{#1}}
\newcommand{\hp}[1]{{#1}}
\newcommand{\hg}[1]{{#1}}
\def\cC{\mathcal{C}}
\def\bsp#1\esp{\begin{split}#1\end{split}}
\newcommand{\refE}[1]{eq.~(\ref{#1})}
\newcommand{\abs}[1]{\left\lvert#1\right\rvert}

\title{Coaction for Feynman integrals and diagrams}

\ShortTitle{Coaction for Feynman integrals and diagrams}

\author{Samuel Abreu\\
       Physikalisches Institut, Albert-Ludwigs-Universit\"at
       Freiburg, D-79104 Freiburg, Germany
      E-mail: \email{samuel.abreu@physik.uni-freiburg.de}}

\author{\speaker{Ruth Britto}
\\
        School of Mathematics and Hamilton Mathematics Institute, Trinity College, Dublin 2, Ireland\\
School of Mathematics, Trinity College, Dublin 2,
        Ireland ;\\
		Institut de Physique Th\'eorique,
		Universit\'e Paris Saclay,
		CEA, CNRS, F-91191 Gif-sur-Yvette
		cedex, France\\
        E-mail: \email{britto@maths.tcd.ie}}

\author{Claude Duhr\\
         Theoretical Physics Department, CERN, Geneva,
        Switzerland ;\\
		Center for Cosmology, Particle Physics and
		Phenomenology (CP3),
		Universit\'e Catholique de
		Louvain, 1348 Louvain-La-Neuve, Belgium\\
        E-mail: \email{claude.duhr@cern.ch}}

\author{Einan Gardi\\
        Higgs Centre for Theoretical Physics, School of
        Physics and Astronomy, The University of Edinburgh,
        Edinburgh EH9 3FD, Scotland, UK\\
        E-mail: \email{einan.gardi@ed.ac.uk}}

\author{James Matthew\\
        Higgs Centre for Theoretical Physics, School of
        Physics and Astronomy, The University of Edinburgh,
        Edinburgh EH9 3FD, Scotland, UK\\
        E-mail: \email{james.matthew@ed.ac.uk}}

\abstract{We propose a general coaction for families of integrals appearing in the evaluation of Feynman diagrams, such as multiple polylogarithms and generalized hypergeometric functions. We further conjecture a link between this coaction and graphical operations on Feynman diagrams. At one-loop order, there is a basis of integrals for which this correspondence is fully explicit. We discuss features and present examples of the diagrammatic coaction on two-loop integrals. We also present the coaction for the functions ${}_{p+1}F_p$ and Appell $F_1$. }

\FullConference{Loops and Legs in Quantum Field Theory (LL2018)\\
		29 April 2018 - 04 May 2018\\
		St. Goar, Germany}

\ifx \tadInsert \undefined
	\def\tadInsert [#1]{
		\raisebox{-4mm}{\includegraphics[keepaspectratio=true, width=.8cm]{./diagrams/#1}}
	}
\fi

\ifx \tadInsertLow \undefined
	\def\tadInsertLow [#1]{
		\raisebox{-4.1mm}{\includegraphics[keepaspectratio=true, width=.81cm]{./diagrams/#1}}
	}
\fi

\ifx \bubInsertHigh \undefined
	\def\bubInsertHigh [#1]{
		\raisebox{-2.2mm}{\includegraphics[keepaspectratio=true, width=2cm]{./diagrams/#1}}
	}
\fi

\ifx \bubInsert \undefined
	\def\bubInsert [#1]{
		\raisebox{-3.6mm}{\includegraphics[keepaspectratio=true, width=2cm]{./diagrams/#1}}
	}
\fi

\ifx \bubInsertLow \undefined
	\def\bubInsertLow [#1]{
		\raisebox{-4.2mm}{\includegraphics[keepaspectratio=true, width=2cm]{./diagrams/#1}}
	}
\fi

\ifx \triInsert \undefined
	\def\triInsert [#1]{
		\raisebox{-6.3mm}{\includegraphics[keepaspectratio=true, width=2cm]{./diagrams/#1}}
	}
\fi

\ifx \triInsertLow \undefined
	\def\triInsertLow [#1]{
		\raisebox{-6.9mm}{\includegraphics[keepaspectratio=true, width=2cm]{./diagrams/#1}}
	}
\fi

\ifx \boxInsert \undefined
	\def\boxInsert [#1]{
		\raisebox{-4.9mm}{\includegraphics[keepaspectratio=true, width=2.3cm]{./diagrams/#1}}
	}
\fi

\ifx \boxInsertLow \undefined
	\def\boxInsertLow [#1]{
		\raisebox{-7.45mm}{\includegraphics[keepaspectratio=true, width=2.3cm]{./diagrams/#1}}
	}
\fi

\begin{document}

% SECTION 1
\section{Introduction}

The class of iterated integrals known as multiple polylogarithms (MPLs) includes many of the functions obtained from integrating Feynman diagrams in dimensional regularization. Their remarkable mathematical properties, including a coaction \cite{gonchCoaction,brownPeriods}, have led to new perspectives on evaluating Feynman diagrams. In particular, the coaction on MPLs has natural links to discontinuities and differential operators, both of which have been important tools in computing scattering processes of interest. 

It is conjectured that the coaction on MPLs corresponds to a similar combinatorial coaction on Feynman diagrams themselves, and that these two coactions agree when the diagrams are evaluated \cite{diagPRL, diagJHEP}. At one-loop order, the conjecture is precise, and there is ample evidence for its validity. 

In this article, we review the conjecture of the diagrammatic coaction and present some progress towards generalizing it beyond one loop on the diagrammatic side, and beyond the class of MPLs to generalized hypergeometric functions on the functional side, eliminating the need to expand in the parameter of dimensional regularization.

The article is structured as follows. We first review the definition of a coaction on a bialgebra. We present the combinatorial incidence coaction, which can be interpreted as the basis of the separate coactions on one-loop Feynman diagrams and MPLs. Then, we state the conjecture of the diagrammatic coaction for Feynman diagrams at one-loop order. We close with new results on the coactions of certain (generalized) hypergeometric and Appell functions, and two-loop  diagrammatic coactions for a double triangle integral and the sunset integral with one massive propagator.

% SECTION 2
\section{Bialgebras and the coaction}

\paragraph{Definitions.} An {\em algebra} $H$ is a ring with a multiplicative unit (1), which is also a vector space over a field $K$. 
In our applications to Feynman integrals, we will always take the field to be $K = \IQ$. 
A {\em bialgebra} is an algebra $H$ with two maps, the {\em coproduct} $\Delta:H\to H\otimes H$, and the {counit} $\varepsilon: H\to\mathbb{Q}$, which are algebra homomorphisms satisfying the axioms $
(\Delta\otimes \textrm{id})\Delta = (\textrm{id}\otimes \Delta)\Delta
$
and $
(\varepsilon\otimes \textrm{id})\Delta = (\textrm{id}\otimes \varepsilon)\Delta = \textrm{id}
$, where id is the identity map. 

\paragraph{The incidence algebra.}
As an exemplar of a bialgebra, let us take the {\em incidence algebra} of \cite{jonirota}, which is a simple combinatorial construction. Let $[n]=\{1,2,\ldots,n\}.$
The elements of the incidence algebra are pairs of nested subsets $[S,T]$, where $S \subseteq T \subseteq [n].$ 
Multiplication is a free abelian operation, and the coproduct is defined by
\begin{eqnarray}
\Delta_{\rm Inc}([S,T]) = \sum_{S \subseteq \hb{X} \subseteq T} [S,X] \otimes [X,T].
\end{eqnarray}

For example:
\begin{eqnarray}
\Delta_{\rm Inc}([S,S]) &=& [S,S]\otimes[S,S] \quad \textrm{for any $S$} \,,
\label{eq:inc-group}
\\
\Delta_{\rm Inc}([\emptyset,\{2\}]) &=& [\emptyset,\hp{\{2\}}] \otimes [\hp{\{2\}},\{2\}] 
+[\emptyset,\hp{\emptyset}] \otimes [\hp{\emptyset},\{2\}] \,,
\label{eq:inc-null-2}
\\
\Delta_{\rm Inc}([\emptyset,\{1,2\}]) &=& [\emptyset,\hp{\{1,2\}}] \otimes [\hp{\{1,2\}},\{1,2\}]
 + [\emptyset,\hp{\{1\}}] \otimes [\hp{\{1\}},\{1,2\}] \nonumber  \\
&& + [\emptyset,\hp{\{2\}}] \otimes [\hp{\{2\}},\{1,2\}] + [\emptyset,\hp{\emptyset}] \otimes [\hp{\emptyset},\{1,2\}]\,,
\label{eq:inc-null-12}
\\
\Delta_{\rm Inc}([\{1\},\{1,2\}]) &=& [\{1\},\hp{\{1,2\}}] \otimes [\hp{\{1,2\}},\{1,2\}] + [\{1\},\hp{\{1\}}] \otimes [\hp{\{1\}},\{1,2\}]  \,,
\label{eq:inc-1-12}
\end{eqnarray}

The counit  $\varepsilon_{\rm Inc}([S,T])$ is 1 if $S=T$, and 0 otherwise.

These combinatorial operations can be applied to edge-sets of graphs. Consider graphs with the topologies of one-loop Feynman diagrams. The set of edges of a graph $G$ is denoted by $E_G$.  In order to include ``cut'' graphs, we let some subset $C$ of $E_G$ carry ``cut'' labels. The cut graph dressed with these labels is then denoted by $(G,C)$. A graph derived from $G$ by contracting edges is denoted $G_X$, where $X$ is the set of edges remaining uncontracted. The empty graph is set to zero identically. Then the incidence coproduct is
\begin{eqnarray}
\label{eq:incGraph}
\Delta_{\rm Inc}(G,C)= \sum_{\substack{C\subseteq X\subseteq E_G\\ X\neq \emptyset}}(G_X,C)\otimes  (G,X)\,.
\end{eqnarray}
Here are the graphical versions of some of the examples listed above, respectively in eqs. (\ref{eq:inc-null-12}), (\ref{eq:inc-1-12}),  and (\ref{eq:inc-group}) with $S=\{1,2\}$.
\begin{eqnarray*}
    \Delta_{\rm Inc}\left[\bubInsertLow[bub2m12Edges]\right]
    &=&\bubInsertLow[bub2m12Edges]\otimes\bubInsert[bub2m12CutPEdges]
    +\tadInsert[tad1]\otimes\bubInsert[bub2m12Cut1Edges]+\tadInsert[tad2]\otimes\bubInsert[bub2m12Cut2Edges]\,, \\
    \Delta_{\rm Inc}\left[\bubInsertLow[bub2m12Cut1Edges]\right]
    &=&\bubInsertLow[bub2m12Cut1Edges]\otimes\bubInsert[bub2m12CutPEdges]
    +\tadInsert[tad1CutM]\otimes\bubInsert[bub2m12Cut1Edges]\,,\\
    \Delta_{\rm Inc}\left[\bubInsertLow[bub2m12CutPEdges]\right]
    &=&\bubInsertLow[bub2m12CutPEdges]\otimes\bubInsert[bub2m12CutPEdges] \,. 
\end{eqnarray*}

\paragraph{A remark on coproduct and coaction.}

If $H$ is a bialgebra over  the field $K$ with coproduct $\Delta:H\to H\otimes H$ and counit $\epsilon:H\to K$, then  a vector space $A$ is called an $H$ (right-) comodule if there is a map $\rho:A\to A\otimes H$ such that $(\rho\otimes \textrm{id})\rho = ( \textrm{id} \otimes \Delta)\rho$ and $( \textrm{id} \otimes \varepsilon )\rho = \textrm{id}$. Then the map $\rho$ is said to be a {\em coaction} on $A$.

The first and second entries in the coaction are thus elements of different mathematical spaces. In the coaction on MPLs, numbers such as $\pi$ can appear only in the first entries. In the corresponding coaction on Feynman diagrams, uncut diagrams (generically) appear only in the first entries.

\paragraph{Coaction on MPLs.}
Multiple polylogarithms (MPLs) are the iterated integrals defined by the following construction. 
\begin{eqnarray}
 G(a_1,a_2,\ldots,a_n;z)=\,\int_0^z\,\frac{d t}{t-a_1}\,G(a_2,\ldots,a_n;t)\,.
 \end{eqnarray}
There is a  coaction on MPLs, graded by their transcendental weight $n$ \cite{gonchCoaction,brownPeriods}.\footnote{With respect to the definition of coaction given above,  $H$ is the bialgebra of MPLs modulo $i\pi$, and $A$ is $Q[i\pi]\otimes H$. } It can be presented as a pairing of contours and integrands, in the same spirit as the incidence algebra.
\begin{eqnarray}
\Delta_{\rm MPL}(G(\vec a;z)) = \sum_{\hb{\vec b} \subseteq \vec a} G(\hb{\vec b};z)\otimes G_{\hb{\vec b}}(\vec a;z)\,.
\end{eqnarray}
On the right-hand side, the subscript ${\vec b}$ means that the integration contour has been deformed such that it encircles each of the points in ${\vec b}$.

Discontinuity and differential operators interact with the coaction in a very simple way, as
\begin{equation}\label{eq:discAndDiff}
	\Delta_{\rm MPL}\textrm{Disc}=(\textrm{Disc}\otimes\textrm{id})\Delta_{\rm MPL}
	\qquad\qquad\textrm{and}\qquad\qquad
	\Delta_{\rm MPL}\partial=(\textrm{id}\otimes\partial)\Delta_{\rm MPL}\,.
\end{equation}
 Because of the grading by weight, it follows that the effects of  discontinuities and differential operators on iterated integrals can be traced to their effects on functions of lower weight.

\paragraph{A remark on the Hopf algebra.}

It is well known that the bialgebra on MPLs can be extended to a Hopf algebra by identifying an antipode map that satisfies the usual axioms, and by working modulo $i\pi$. To extend the incidence algebra to a Hopf algebra, it is necessary to adjoin formal multiplicative inverses of the ``grouplike'' elements $[S,S]$. The existence of grouplike elements means that the incidence Hopf algebra is not connected. It remains to be determined whether the antipode map associated to our diagrammatic coaction might carry physical significance. We note that in our diagrammatic coaction, maximally cut one-loop integrals are naturally appearing grouplike elements. They evaluate to functions of the form $x^\eps$, whose infinite Laurent series in MPLs is grouplike, even though there are no individuallly grouplike MPLs.

% SECTION 3
\section{Diagrammatic coaction for Feynman integrals}

We now outline the diagrammatic coaction, by describing how to interpret one-loop diagrams with Feynman rules, and by deforming the incidence coaction on graphs so that it matches the coaction on MPLs. Some early applications may be found in  \cite{diagJHEP,Abreu:2018sat}. We then recast the diagrammatic coaction as a special case of a general coaction on families of integrals.

\paragraph{Interpretation of diagrams.} It is possible to reduce one-loop Feynman integrals to the following scalar basis corresponding to graphs $G$ with edges $E_G$:
\begin{eqnarray}
\label{eq:defJG}
J_G &=& \hg{\frac{i e^{\gamma_E\eps}}{\pi^{\abs{E_G}/2}}} \int d^{D_G} k\prod_{j \in E_G}\frac{1}{(k-q_j)^2-m_j^2}
\equiv \int_{\Gamma_\emptyset} \omega_G\,,
\end{eqnarray}
where  $k$ is the loop momentum, $q_j$ are sums of external momenta, $m_j$ are internal masses, and the integrals are evaluated in the dimensionality $D_G \equiv D_{\abs{E_G}}$ given as follows.
\begin{equation}
\label{eq:defDn}
D_n=\left\{\begin{array}{ll}
n-2\eps\,, & \textrm{ for } n \textrm{ even}\,,\\
n+1-2\eps\,, & \textrm{ for } n \textrm{ odd}\,.
\end{array}\right.
\end{equation}
Each of the $J_G$ is of uniform transcendental weight when expanded in  $\eps$, if we assign a weight of -1 to $\eps$ itself.

We also need to define the cuts of  Feynman integrals. This is done using residues, so that we deform the integration contour instead of inserting on-shell delta functions in the integrand. Our definition is written as 
\begin{eqnarray}
\label{eq:defCJG}
\cC_C[J_G] 
= \int_{\Gamma_C} \omega_G  \, \,\mod i\pi\,.
\end{eqnarray}
Here, $C$ denotes the set of cut propagators, and the contour $\Gamma_C$ is defined to encircle the poles associated to $C$, in contrast to the original contour $\Gamma_\emptyset$ of \refE{eq:defJG}. This information is sufficient to determine the cut integral up to terms proportional to $i\pi$, which are eliminated by a quotient construction, and up to an overall sign. The freedom in this definition is consistent with the fact that the coaction is blind to such terms in the second entry. Further details of the definition, along with discussion and examples, may be found in \cite{oneLoopCuts}.

\paragraph{Deformation of the incidence coaction.}
An algebraic isomorphism of the incidence bialgebra leads to the following deformed version,
\begin{equation}\bsp
\label{eq:phipsiaction}
\Delta_{a} (G,C)
&\,=\sum_{\substack{C\subseteq X\subseteq E_G,\\X\neq \emptyset}} \left((G_X,C) +a_X \sum_{e\in X\setminus C} (G_{X\setminus e},C) \right) \otimes (G,X)\,.
\esp\end{equation}
where $a_X=a$ if $|X|$ is even and 0 otherwise, and $a$ is a constant.
 The value $a=0$ reproduces the undeformed coaction, \refE{eq:incGraph}. In order to match the coactions on one-loop graphs and their expressions in MPLs, we will need the particular deformation
with $a=1/2$.
This value of $a$ and the dependence of $a_X$ on the parity of $|X|$ have their origin in homology theory, as will be seen below.

\paragraph{Statement of diagrammatic conjecture.}
For one-loop graphs,
\begin{equation}\label{eq:graph_conjecture}
\Delta_{\textrm{MPL}} \, \mathcal{I}(G,C) = (\mathcal{I}\otimes\mathcal{I}) \, \Delta_{1/2}(G,C)\,,
\end{equation}
where  by $\mathcal{I}$ we denote the linear map that associates to $(G,C)$ its cut integral $\mathcal{I}(G,C)\equiv\cC_C {J}_G$ in $D_G$ dimensions, as defined above, and performs a Laurent expansion in $\eps$. 
For example,
\begin{eqnarray}
	\Delta
	\left[\bubInsertLow[bub2m12Edges]\right]
	&=& \bubInsertLow[bub2m12Edges] \otimes \bubInsert[bub2m12CutPEdges] 
	+\tadInsert[tad1]\otimes \left( \bubInsert[bub2m12Cut1Edges] + \half\bubInsert[bub2m12CutPEdges] \right) 
	\nonumber \\
	&&+\tadInsert[tad2]\otimes \left(\bubInsert[bub2m12Cut2Edges] + \half\bubInsert[bub2m12CutPEdges]  \right)\,,
\label{diagConjBubM1M2}
\\
	\Delta\left[\triInsertLow[t1m1LabelEdges]\right] &=&
	\tadInsert[tad1]\otimes \left( \triInsertLow[t1m1Cut1LabelEdges] + \half \triInsertLow[t1m1Cut12LabelEdges]\right) 
	+
	\bubInsertLow[bub1m1LabelEdges] \otimes\triInsertLow[t1m1Cut12LabelEdges]\,,
\label{eq:1loopex2F1} 
	\\
\label{eq:1loopexF1}
	\Delta\left[\triInsertLow[t1m12]\right] &=&
	\tadInsert[tad1]\otimes \left( \triInsertLow[t1m12Cut1] + \half \triInsertLow[t1m12Cut12]\right)  +	\tadInsert[tad2]\otimes \left( \triInsertLow[t1m12Cut2] + \half \triInsertLow[t1m12Cut12]\right)  \nonumber \\
&&	+
	\bubInsertLow[bub2m12Edges] \otimes\triInsertLow[t1m12Cut12] + \triInsertLow[t1m12] \otimes \triInsertLow[t1m12Cut123]\,.
\end{eqnarray}
where the coaction can either be read directly as $\Delta_{1/2}$ acting on the diagrams, or interpreted as $\Delta_{\rm MPL}$ after applying the Feynman rules. 
We observe nontrivial cancellations among terms at different orders in $\eps$ on the right-hand side, resulting in a coaction that is valid order by order.
In the following section, we will extend the coaction to hypergeometric functions including ${}_2F_1$ and Appell $F_1$, eliminating the need to expand in $\eps$.

  Taking the first of these examples as an illustration, \refE{diagConjBubM1M2} can be written in terms of integrals as 
\begin{eqnarray}
\Delta\left(\int_{\Gamma_\emptyset} \omega_{12}\right)	&=& \int_{\Gamma_\emptyset} \omega_{12} \otimes \int_{\Gamma_{12}} \omega_{12} 
+ \int_{\Gamma_\emptyset} \omega_{1} \otimes \left( \int_{\Gamma_{1}} \omega_{12} + \half \int_{\Gamma_{12}} \omega_{12} \right)
+ \int_{\Gamma_\emptyset} \omega_{2} \otimes \left( \int_{\Gamma_{2}} \omega_{12} + \half \int_{\Gamma_{12}} \omega_{12} \right)
\nonumber
 \\
&=& \int_{\Gamma_{\emptyset}} \omega_{\hb{12}} \otimes \int_{\Gamma_{\hb{12}}} \omega_{12} 
+ \int_{\Gamma_\emptyset} \omega_{\hb{1}} \otimes  \int_{-\half\Gamma_{\hb{1\infty}}} \omega_{12}  
+ \int_{\Gamma_\emptyset} \omega_{\hb{2}} \otimes  \int_{-\half\Gamma_{\hb{2\infty}}} \omega_{12}  \,,
\end{eqnarray}
where in the last line, we have introduced  contours detecting the pole at infinite momentum. In fact, it is the pole at infinity that leads to the need for deforming the incidence coaction on graphs. At one loop, the Decomposition Theorem for the homology of Feynman integrals shows how to rewrite any contour involving infinity in terms of the ones describing cuts of  propagators, leading to the equivalence of the two lines above \cite{homology}.

The diagrammatic coaction can be written neatly, for all one-loop integrals, as
\begin{equation}\label{eq:oneLoopMF}
	\Delta\left(\int_{\gamma}\omega_G \right)=\sum_{\emptyset\neq C\subseteq E_G}
	\int_{\gamma}\omega_{G_C}\otimes
	\int_{\gamma_C}\omega_G\,,
\end{equation}
where 
$
\gamma_C
=	\Gamma_C+a_C\sum_{e\in E_G\setminus C}\Gamma_{Ce}\,,
$
and $a_C=1/2$ for $|C|$ odd and 0 otherwise. We note that since $\gamma_\emptyset$ is excluded in the second entry, uncut integrals can appear only in the first entry.
\paragraph{Master formula for coaction on integrals.}
The diagrammatic coaction of \refE{eq:oneLoopMF} is a special case of a general 
 coaction on integrals  defined by 
\begin{equation}\label{eq:masterformula}
	\Delta\left(\int_\gamma\omega\right)=
	\sum_i\int_\gamma\omega_i\otimes
	\int_{\gamma_i}\omega\,,
\end{equation}
whenever there is a suitable pairing of master integrands and master contours such that
\begin{equation}\label{eq:ssProj}
	P_{ss}\left(\int_{\gamma_i}\omega_j\right)=\delta_{ij}\,.
\end{equation}
Here, $P_{ss}$ denotes the projection onto semi-simple numbers, which are numbers $x$ satisfying $\Delta(x) = x \otimes 1.$ In particular, $P_{ss}$ retains $\pi$, but drops polylogarithms unless they evaluate to powers of $\pi$.

% SECTION 4
\section{Coaction on hypergeometric functions}

In this section, we present bases of integrands and integration regions adapted for various (generalized) hypergeometric functions giving coactions of the form 
(\ref{eq:masterformula}), which moreover correspond to the coaction on Feynman diagrams.
Specifically, in each case, $\int_{\gamma_1} (\sum_j a_j \omega_j)$ evaluates to a hypergeometric function in a given family. We note that in each of the cases presented here, there is a simple change of variables such that each integral over $\gamma_i (i>1)$ can be rewritten as an integral over $\gamma_1$, and hence recognized as a hypergeometric integral of the same family. 

We have checked that the coactions written here are consistent with the coaction on MPLs in the $\eps$ expansion.

\paragraph{Hypergeometric ${}_2F_1$.}

Consider the family of integrands
$\omega(\a_1,\a_2,\a_3;z) = x^{\a_1} (1-x)^{\a_2} (1-zx)^{\a_3} dx$. They are related to the Gauss hypergeometric function ${}_2F_1$ by
\begin{eqnarray}
\int_0^1 \omega(\a_1,\a_2,\a_3) = \hg{\frac{\Gamma(\a_1)\Gamma(\a_2-\a_1)}{\Gamma(\a_2)}} \, {}_2F_1(-\a_3,\a_1+1;\a_2+\a_1+2;z)\,.
\end{eqnarray}
Through integration-by-parts relations, it is possible to shift the exponents by integer values and reduce to a set of two master integrands. For the ${}_2F_1$ family, this statement is a consequence of the well-known contiguous relations. If we expand around integer arguments, i.e. $ \a_i = n_i + a_i\eps$ and  $n_i \in \IZ,$ $a_i\eps \not\in \IZ,$ we can take the master integrands to be 
\begin{eqnarray}
\omega_0 = a_2\eps\,\, x^{a_1\eps} (1-x)^{-1+a_2\eps} (1-zx)^{a_3\eps}\,, \qquad
\omega_1 = a_3\eps\, z\,\, x^{a_1\eps} (1-x)^{a_2\eps} (1-zx)^{-1+a_3\eps} \,.
\end{eqnarray}
Normalization factors have been chosen so that with the two contours $\gamma_0=[0,1]$ and $\gamma_1=[0,1/z]$, we have $P_{ss}\int_{\gamma_i}\omega_j = \delta_{ij}$.
Since the diagrams in (\ref{diagConjBubM1M2}) and  (\ref{eq:1loopex2F1}) evaluate to ${}_2F_1$ functions, we can verify that the coaction (\ref{eq:masterformula}) accounts for those diagrammatic formulas, without having to expand in $\eps$.\footnote{One might wish to write the coaction on a ${}_2F_1$ function unaccompanied by the gamma-function prefactors. This can be done with the help of the relation $\Delta(\Gamma(1+a\eps))=\Gamma(1+a\eps) \otimes \Gamma(1+a\eps)$.}

\paragraph{Appell $F_1$.}
For this family, the integrands take the form
$
\omega(\a_1,\a_2,\a_3,\a_4;z_1,z_2) =
x^{\a_1} (1-x)^{\a_2}(1-z_1 x)^{\a_3}(1-z_2 x)^{\a_4}\,dx\,,
$
since 
\begin{eqnarray}
\int_0^1 \omega(\a_1,\a_2,\a_3,\a_4;z_1,z_2) =
 \frac{\Gamma(\a_1)\Gamma(\a_2-\a_1)}{\Gamma(\a_2)} \, F_1(\a_1,\a_3,\a_4,\a_2;z_1,z_2)\,.
\end{eqnarray}
If again we expand around integer arguments, $\a_i = n_i + a_i\eps$ and  $n_i \in \IZ$, $a_i\eps \not\in \IZ,$ then the following choice of master integrands, 
\begin{eqnarray}
\omega_0 &=& a_2\eps\,\,  x^{a_1\eps} (1-x)^{-1+a_2\eps} (1-z_1 x)^{a_3\eps}(1-z_2 x)^{a_4\eps} \,,  \\
\omega_1 &=&  a_3\eps\, z_1\,\, x^{a_1\eps} (1-x)^{a_2\eps} (1-z_1 x)^{-1+a_3\eps}(1-z_2 x)^{a_4\eps}\,,   \\
\omega_2 &=&  a_4\eps\, z_2\,\, x^{a_1\eps} (1-x)^{a_2\eps} (1-z_1 x)^{a_3\eps}(1-z_2 x)^{-1+a_4\eps}  \,,
\end{eqnarray}
with the corresponding master contours $\gamma_0=[0,1]$, $\gamma_1=[0,z_1^{-1}]$, $\gamma_2=[0,z_2^{-1}]$, satisfies
$P_{ss}\int_{\gamma_i}\omega_j = \delta_{ij}$. 
Since the diagrams in (\ref{eq:1loopexF1}) evaluate to $F_1$ functions, we can verify that the coaction (\ref{eq:masterformula}) accounts for the diagrammatic formula, without having to expand in $\eps$.

\paragraph{Generalized hypergeometric ${}_{p+1}F_p$.}
We show the case $p=2$, from which it is straightforward to  generalize to arbitrary $p$.
The integrands are
$
\omega(\a_1,\a_2,\a_3,\a_4,\a_5;z) =
x^{\a_1} (1-x)^{\a_2}
y^{\a_3} (1-y)^{\a_4}  
(1-zxy)^{\a_5} \,dx\,dy
$,
 where $\a_i = n_i + a_i\eps$ and  $n_i \in \IZ,$ $a_i\eps \not\in \IZ,$
Then 
\begin{eqnarray}
&&
 \int_0^1 \int_0^1 \omega(\a_1,\a_2,\a_3,\a_4,\a_5;z) 
=
\\&&
 \frac{\Gamma(\a_1+1)\Gamma(\a_2+1)\Gamma(\a_3+1)\Gamma(\a_4+1)}{\Gamma(2+\a_1+\a_2)\Gamma(2+\a_3+\a_4)} \, {}_3F_2(\a_1+1,\a_3+1,-\a_5;2+\a_1+\a_2,2+\a_3+\a_4;z)\,. \nonumber
\end{eqnarray}
With the master integrands
\begin{eqnarray}
\omega_0 &=& a_2 a_4 \,\eps^2 x^{a_1\eps} (1-x)^{-1+a_2\eps} y^{a_3\eps} (1-y)^{-1+a_4\eps}(1-zxy)^{a_5\eps}\,, \\
\omega_1 &=&  \frac{a_2 a_5(a_1-a_3-a_4)}{a_1+a_2-a_3-a_4} 
\eps^2 z\, x^{a_1\eps} (1-x)^{-1+a_2\eps} y^{a_3\eps} (1-y)^{a_4\eps}(1-zxy)^{-1+a_5\eps}\,, \\
\omega_2 &=& 
\frac{a_4 a_5(a_3-a_1-a_2)}{a_3+a_4-a_1-a_2}
\eps^2 z \,
 x^{a_1\eps} (1-x)^{a_2\eps} y^{a_3\eps} (1-y)^{-1+a_4\eps}(1-zxy)^{-1+a_5\eps} \,.
\end{eqnarray}
and  contours $\gamma_0=\int_0^1 dx \int_0^1 dy$, $\gamma_1=\int_0^1 dx \int_0^{1/zx} dy$, $\gamma_2=\int_0^1 dy \int_0^{1/zy} dx$, 
we have $P_{ss}\int_{\gamma_i}\omega_j = \delta_{ij}$.

In the following diagrammatic relation, the diagram on the left-hand-side evalutes to a ${}_{3}F_2$ function. 
For this function, the coaction above is consistent with a surprisingly direct application of the incidence coaction on diagrams.
\begin{eqnarray*}
&&	\Delta\left[\boxInsert[DTri12]\right]= 
\boxInsert[DTri12] \otimes \boxInsert[DTri12C12345] 
+ \bubInsert[sunrise0m1] \otimes \boxInsert[DTri12C234] \\
&& 
+ \bubInsert[sunrise0m2] \otimes \boxInsert[DTri12C135] 
+ \boxInsert[PHat12] \otimes \boxInsert[DTri12C1345] \\
&& 
+ \boxInsert[PHat21] \otimes \boxInsert[DTri12C2345]
+ \bubInsertHigh[DBub12] \otimes \boxInsert[DTri12C1245]
\end{eqnarray*}
The graphs in this equation are interpreted as scalar integrals in $4-2\eps$ dimensions, adjusted by simple rational factors in order to convert them to pure integrals. Specifically, these factors are $(1-2\eps)(1-3\eps)(2-3\eps)/p^2$ for the sunset, $(1-2\eps)(1-3\eps)$ for the triangle-bubble, and $(1-2\eps)^2$ for the double bubble. 

Further details of the results in this section will appear in forthcoming publications \cite{abdgw}.

%SECTION 5
\section{Discussion}

In attempting to generalize the diagrammatic coaction beyond one-loop, we further expect to find a matrix equation for all master integrals of any given topology. Consider, for example, the sunset integral with external momentum $p$ and just one massive propagator of mass $m$. There are two master integrals in the top topology, which can be taken to be the sunset with single and double powers, respectively, of the massive propagator:
\begin{eqnarray}
S_{111}  = \int_{\Gamma_\emptyset}\omega_{111} =
 \bubInsert[sunrise1m]\,,
\qquad
S_{211} 
= \int_{\Gamma_\emptyset}\omega_{121} = \bubInsert[sunrise1mSqP]\,.
\end{eqnarray}
For each master integral, we find that only two of the cuts are linearly independent, in keeping with the arguments of \cite{kaspersimon}. If we label the massive propagator by 1 and the others by 2 and 3, the  integration contours for cuts are related as follows:
$
\Gamma_{12} = \Gamma_{13} ,
0 = 4\Gamma_{23}+3\Gamma_{1}-2\Gamma_{12}+\Gamma_{123} ,
0 = \Gamma_{1}-\Gamma_{12}+\Gamma_{23} ,
\eps \Gamma_{\emptyset} = {\Gamma(1+\eps)\Gamma(1-\eps)} \Gamma_{23}
$.
Because of their relation to the physical discontinuities of the sunset integral, let us
 choose the contours $\Gamma_{1},\Gamma_{123}$ as a basis, giving integrals such as 
\begin{eqnarray}
\cC_{1}S_{111}=\int_{\Gamma_1}\omega_{111}= \bubInsert[sunrise1mC1], 
\qquad
\cC_{123}S_{111}=\int_{\Gamma_{123}}\omega_{111}=\bubInsert[sunrise1mC123]\,.
\end{eqnarray}
We find that the coaction on $S_{111}$, as defined for the ${}_2F_1$ family, can be written as
a linear combination of terms with $S_{111}$ and $S_{211}$ in the first entries, and $\cC_{1}S_{111}$ and $\cC_{123}S_{111}$ in the second entries \cite{abdgw}. From this relation, we can confirm that $\cC_{1}S_{111}$ and $\cC_{123}S_{111}$ are precisely the discontinuities related to the weight-one first entries $\log(m^2)$ and $\log(p^2-m^2)$, as expected. The coaction on $S_{211}$ is similar.
We emphasize the form of the result: the coaction on the integral can be expressed in terms of Feynman diagrams, and the second entries are generically cut integrals.

Similarly to the one-loop case, we observe that it is helpful to identify a basis of pure integrals. In this example, $\omega_1 \equiv \eps (p^2-m^2)\omega_{111}$ yields a pure integral when paired with any of the cut or uncut contours. For the second master integral, it is possible to subtract an algebraic multiple of $\omega_1$ from $m^2(p^2-m^2) \omega_{211}$ to leave a pure integrand $\omega_2$. We can now neatly present the full coaction for this topology. Construct the matrix 
\begin{eqnarray}
\Omega = \left(
\begin{array}{cc}
\int_{\Gamma_{\emptyset}} \omega_1 &  \int_{\Gamma_{\emptyset}} \omega_2 \\
\int_{-\frac{1}{3}\Gamma_{1}-\frac{1}{2}\Gamma_{123}} \omega_1 &  \int_{-\frac{1}{3}\Gamma_{1}-\frac{1}{2}\Gamma_{123}} \omega_2 \\
\end{array}
\right)\,,
\end{eqnarray}
where the contours and normalizations have been chosen such that $P_{ss}\Omega$ is the identity matrix, and the first row contains the original master integrals.
Then the coaction on any integral in this family can be obtained with the formula $\Delta(\Omega_{ij})=\sum_k \Omega_{ik} \otimes \Omega_{kj}.$

In view of our results, it would be very interesting to seek a systematic diagrammatic coaction for multiloop Feynman integrals and amplitudes, supported by a coaction on an expanded set of functions including various generalizations of hypergeometric functions.

\acknowledgments{We thank D. Broadhurst, D. Kreimer, and E. Panzer for helpful feedback on the conference presentation, and the organizers of the conference.
This work is supported by
the Alexander von Humboldt Foundation
in the framework of the Sofja Kovalevskaja Award 2014, endowed
by the German Federal Ministry of Education and Research (SA),
the ERC Consolidator Grant 647356 ``CutLoops'' (RB), the ERC 
Starting Grant 637019 ``MathAm'' (CD), and the STFC 
Consolidated Grant
``Particle Physics at the Higgs Centre''~(EG, JM).
}


\begin{thebibliography}{99}

\bibitem{gonchCoaction}
  A.~B.~Goncharov,
  arXiv:math/0208144.

\bibitem{brownPeriods}
  F.~Brown,
  arXiv:1512.06410 [math.NT].


\bibitem{diagPRL}
  S.~Abreu, R.~Britto, C.~Duhr and E.~Gardi,
  Phys.\ Rev.\ Lett.\  {\bf 119} (2017) no.5,  051601,
  arXiv:1703.05064 [hep-th].

\bibitem{diagJHEP}
  S.~Abreu, R.~Britto, C.~Duhr and E.~Gardi,
  JHEP {\bf 1712} (2017) 090,
  arXiv:1704.07931 [hep-th].
  
\bibitem{Abreu:2018sat} 
  S.~Abreu, R.~Britto, C.~Duhr and E.~Gardi,
  PoS RADCOR {\bf 2017}, 002 (2018)
  doi:10.22323/1.290.0002
  [arXiv:1803.05894 [hep-th]].

\bibitem{jonirota}
S.~A.~Joni and G.~C.~Rota, Stud.\ Appl.\ Math.\ {\bf 61} (1979) 93.

\bibitem{oneLoopCuts}
  S.~Abreu, R.~Britto, C.~Duhr and E.~Gardi,
  JHEP {\bf 1706} (2017) 114,
  arXiv:1702.03163 [hep-th].

\bibitem{homology}
  D.~Fotiadi, M.~Froissart, J.~Lascoux, F.~Pham,
  Topology {\bf 4},
  159-191,
  Pergamon Press,
  (1965); 
  R.~C.~Hwa and V.~L.~Teplitz,
  \emph{Homology and Feynman integrals}, 
  W.~A.~Benjamin Inc.,
  (1966).


\bibitem{abdgw}
  S.~Abreu, R.~Britto, C.~Duhr, E.~Gardi and J.~Matthew, in preparation.


\bibitem{kaspersimon}
  S.~Caron-Huot and K.~J.~Larsen,
  JHEP {\bf 1210}, 026 (2012)
  doi:10.1007/JHEP10(2012)026
  [arXiv:1205.0801 [hep-ph]].





\end{thebibliography}
\end{document}